\documentclass[letterpaper, 10pt]{IEEEtran}
\usepackage[top=0.75in, bottom=0.75in, left=0.625in, right=0.625in]{geometry}
\IEEEoverridecommandlockouts 
\usepackage{amsmath}
\usepackage{amsthm}
\usepackage{amsfonts}
\usepackage{amssymb}
\usepackage{hyperref}
\usepackage{array}                      
\usepackage{cite}
\usepackage{mathtools}
\usepackage{comment}
\usepackage{xcolor}
\usepackage{mathrsfs}
\usepackage{caption}
\usepackage{subcaption}
\usepackage{tabularx}
\usepackage{graphicx}

\usepackage[font=small,labelfont=bf]{caption}
\usepackage [english]{babel}
\usepackage [autostyle, english = american]{csquotes}
\MakeOuterQuote{"}
\usepackage[compact]{titlesec}

\newcommand{\RR}{\mathbb{R}}
\newcommand{\ZZ}{\mathbb{Z}}
\newcommand{\dd}{{\tt d}}
\newcommand{\ball}{\mathbb{B}}

\newcommand{\diag}[1]{\text{diag}(#1)}

\newcommand{\Span}[1]{\text{Span}(#1)}
\newcommand{\Range}[1]{\text{Range}(#1)}

\newcommand{\verts}{\mathcal{V}}
\newcommand{\edges}{\mathcal{E}}
\newcommand{\graph}{\mathcal{G}}
\newcommand{\nbd}{\mathcal{N}}
\newcommand{\adj}{\mathsf{A}}
\newcommand{\lap}{\mathsf{L}}

\newcommand{\dom}{\text{dom }}
\newcommand{\hyb}{\mathcal{H}}
\newcommand{\attr}{\mathcal{A}}
\newcommand{\statespc}{\mathcal{X}}

\newcommand{\sproj}{\mathsf{S}}
\newcommand{\WV}{\mathsf{V}}
\newcommand{\WD}{\mathsf{D}}
\newcommand{\mmat}{\mathsf{M}}
\newcommand{\fmat}{\mathsf{F}}
\newcommand{\tspc}{\mathcal{T}}
\newcommand{\agreement}{\mathfrak{A}}
\newcommand{\TT}{\mathsf{T}}

\newcommand{\update}[1]{\textcolor{black}{#1}}

\theoremstyle{definition}
\newtheorem{theorem}{Theorem}

\newtheorem{corollary}{Corollary}
             
\newtheorem{definition}{Definition}

\newtheorem{lemma}{Lemma}

\begin{document}
\pagenumbering{gobble}
\title{ChronoSync: A Decentralized Chronometer Synchronization Protocol for Multi-Agent Systems}

\author{Federico M. Zegers and Sean Phillips
\thanks{F. M. Zegers is with \update{the} Johns Hopkins University Applied Physics Laboratory,
Laurel, MD 20723, USA. E-mail: {\tt federico.zegers}@{\tt jhuapl.edu}.}
\thanks{S. Phillips is with the Air Force Research Laboratory, Space Vehicles
Directorate, Kirtland AFB, NM 87117, USA. Approved for public release;
distribution unlimited. Public Affairs approved AFRL-2025-1776.}
}

\vspace{0.25in}
\maketitle
\thispagestyle{empty}
\begin{abstract}
This work presents a decentralized time synchronization algorithm for multi-agent systems. 
Each agent possesses two clocks, a hardware clock that is perturbed by environmental phenomena (e.g., temperature, humidity, pressure, $g$ forces, etc.) and a steerable software clock that inherits the perturbations affecting the hardware clock.
Under these disturbances and the independent time kept by the hardware clocks, our consensus-based controller enables all agents to steer their software-defined clocks into practical synchronization while achieving a common user-defined clock drift.
Furthermore, we treat the drift of each hardware clock as an unknown parameter, which our algorithm can accurately estimate.
The coupling of the agents is modeled by a connected, undirected, and static graph.
However, each agent possesses a timer mechanism that determines when to broadcast a sample of its software time and update its own software-time estimate.
Hence, communication between agents can be directed, intermittent, and asynchronous.
The closed-loop dynamics of the ensemble is modeled using a hybrid system, where a Lyapunov-based stability analysis demonstrates that a set encoding the time synchronization and clock drift estimation objectives is globally practically exponentially stable.
The performance suggested by the theoretical development is confirmed in simulation.
\end{abstract}

\section{Introduction}
\subsection{Motivation}
Much of the recent research conducted on multi-agent systems (MASs) focuses on variations of the consensus problem, which enable capabilities such as trajectory synchronization, rendezvous, formation control, distributed state estimation, and time synchronization~\cite{AKYILDIZ2002393}.
The tools developed by this body of work facilitate the study of sensor networks, cellular networks, satellite systems, and autonomous driving vehicles~\cite{AKYILDIZ2002393, zheng2009wireless}.
As the world has become more connected through the proliferation of embedded systems, the time-synchronization problem has increased in interest.
Time synchronization touches nearly every facet of modern technology, and accurate time synchronization is essential for things like autonomous safety systems based on reachability analysis, propagating state estimates using various filter types, and task scheduling between multiple \update{assets---think of} airliner traffic flowing through a major airport.
Today, time synchronization is achieved \update{by utilizing} satellite constellations dedicated to position, navigation, and timing (PNT), \update{which} are coupled with ground-based atomic clock standards.
Three prominent examples of \update{PNT satellite} constellations are NavStar Global Positioning System, Global'naya Navigatsionnaya Sputnikovaya Sistema (\update{i.e.,} GLONASS), and the BeiDou Navigation Satellite System.
Yet, as the global radio frequency spectrum becomes more congested due to increased use and signals from PNT constellations become \update{faint} as a result of increased radio traffic, access to accurate time relative to a global time standard becomes uncertain, which may have a far-reaching impact. 

\subsection{Literature Review}
On the face of it, clocks are simple.
An oscillator functions as a fixed-frequency generator and a counter tallies the number of cycles completed by the oscillator.
One second corresponds to the completion of $k$ cycles by the oscillator for some fixed positive integer $k$.
However, each clock experiences perturbations which may cause variations in frequency generation and cycle tallying.
Although these variations may be minuscule, they have a compounding effect.
The time kept by two initially synchronized clocks will drift apart as a result of variations in frequency generation and cycle tallying if not corrected.
Given the importance of time synchronization, many researchers have devised methods of steering clocks into agreement.

One of the most influential results within the time synchronization literature is the so-called Network Time Protocol~\cite{Mills1991}, a centralized algorithm wherein a group of follower agents synchronize their software-defined times with that of a leader agent while accounting for communication delays.
Other centralized algorithms include~\cite{Elson.Girod.ea2003,Ganeriwal.Kumar.ea2003}.
To assuage the limitations resulting from centralization, researchers developed distributed consensus-based time synchronization protocols, e.g., \cite{Garone.Gasparri.ea2015,Kikuya.Dibaji.ea2018,Guarro.Sanfelice2023}.
The result in~\cite{Guarro.Sanfelice2023} is of particular significance since it provides a Lyapunov-based stability analysis that renders a set describing the time synchronization objective globally exponentially stable \update{(GES)} for the corresponding hybrid system model.
From this stability analysis \update{stems} robustness properties that provide global performance guarantees despite the existence of perturbations influencing information broadcasts, causing communication delays, and generating variations in clock drift (\update{that is}, a clock's time rate of change)\update{---these} robustness properties are derived via input-to-state stability arguments. 

\subsection{Contribution}
Inspired by~\cite{Guarro.Sanfelice2023}, this paper presents \texttt{ChronoSync}, a novel decentralized consensus-based protocol for MASs facilitating the synchronization of software-defined times.
Similar to the hybrid network time protocol (HyNTP) of~\cite{Guarro.Sanfelice2023}, \texttt{ChronoSync} is distributed since agents only require access to information from $1$-hop neighbors and can make decisions independently to achieve the global objective of time synchronization in the software-defined times.
Furthermore, both algorithms employ software-time samples obtained from intermittent communication.
However, unlike HyNTP, \texttt{ChronoSync} is decentralized and accommodates the asynchronous exchange of information between neighboring agents while also possessing global stability and robustness properties.
Of course, these properties belong to an attractor-hybrid system \update{pairing} that differ from \update{that} in~\cite{Guarro.Sanfelice2023}.
\update{Using} the framework in~\cite{Goebel.Sanfelice.ea2012}, we provide a hybrid system model for the ensemble system, transform the software-time synchronization problem into a set stabilization problem, and show \update{that} the desired attractor is globally practically exponentially stable\footnote{\update{Although formally defined later, a set $\attr$ is globally practically exponentially stable for a hybrid system $\hyb$ if all maximal solutions of $\hyb$ are complete and converge exponentially to a closed superset of $\attr$.}} via a Lyapunov analysis in the presence of perturbations---the attractor is \update{GES} in the absence of perturbations.
To demonstrate the performance of our protocol, results from a simulation are provided towards the end of this paper.
%

\section{Preliminaries}\label{sec: Preliminaries}
\subsection{Notation}
Given a constant $a\in\RR$, let $\RR_{\geq a}\coloneqq[a,\infty)$,
$\RR_{>a}\coloneqq(a,\infty)$, $\ZZ_{\geq a}\coloneqq\RR_{\geq a}\cap\ZZ$, and $\ZZ_{>a}\coloneqq\RR_{>a}\cap\ZZ$.
For $p,q\in\ZZ_{>0}$, the $p\times q$
zero matrix and the $p\times 1$ zero column vector are respectively denoted by
$0_{p\times q}$ and $0_p$.
When it is inconvenient to specify the dimension of a zero matrix or vector, we will write $\mathbf{0}$.
The $p\times p$ identity
matrix and the $p\times 1$ column vector with all entries being one are denoted by $I_p$ and $1_p$, respectively. 
The Euclidean norm of $r\in\RR^p$
is denoted by $\Vert r\Vert \coloneqq\sqrt{r^\top r}$.
For $M\in\ZZ_{\geq 2}$, let $[M]\coloneqq\{1,2,...,M\}$.
The maximum and minimum eigenvalues of a real symmetric matrix $A\in\RR^{n\times n}$ are denoted by $\lambda_{\max}(A)$ and $\lambda_{\min}(A)$, respectively.
The block diagonal matrix with general blocks $G_1,G_2,...,G_p$ is denoted by $\text{diag}(G_1,G_2,...,G_p)$.
The distance of a point $r\in\RR^p$ to the set $S\subset\RR^p$ is given by $\vert r\vert_{S}\coloneqq\inf\{\Vert r-s\Vert\colon s\in S\}\in\RR_{\geq 0}$.
Furthermore, let $r+S\coloneqq \{r+s\in\RR^p\colon s\in S\}$, and, for any matrix $K\in\RR^{n\times p}$, let $KS\coloneqq\{Ks\in\RR^n\colon s\in S \}$.
The cartesian product of $S_1$ and $S_2$ is denoted by $S_1\times S_2$.
Let $\ball\coloneqq [-1,1]$.
Given a collection of vectors $\{z_1,z_2,...,z_p\}\subset\RR^q$, let $(z_k)_{k\in[p]}\coloneqq [z_1^\top,z_2^\top,...,z_p^\top]^\top\in\RR^{pq}$.
Similarly, for $x\in\RR^p$ and $y\in\RR^q$, let $(x,y)\coloneqq [x^\top,y^\top]^\top\in\RR^{p+q}$.
For any nonempty sets $A$ and $B$, the single-valued map $f$ and the set-valued map $F$ with domain $A$ and codomain $B$ are denoted by $f\colon A\to B$ and $F\colon A\rightrightarrows B$, respectively.
The set-valued derivative of a continuously differentiable function $h\colon\RR^n\to\RR$ with respect to the differential inclusion $\dot{x}\in F(x)$ such that $F\colon\RR^n\rightrightarrows\RR^n$ is denoted by $\{\nabla h(x)^\top v\colon v\in F(x)\}$.
The analysis carried out in this work is based on the hybrid systems framework developed in~\cite{Goebel.Sanfelice.ea2012}.
Please consult this reference for questions regarding notation not defined here.

\subsection{Graphs}
Let $\graph\coloneqq(\verts,\edges)$ be a graph on $N\in\ZZ_{\geq2}$ nodes, where $\verts\coloneqq[N]$ denotes the node set and $\edges\subseteq\verts\times\verts$ denotes the edge set.
If $(p,q)\in\edges$ implies $(q,p)\in\edges$ for all distinct nodes $p,q\in\verts$, then the graph $\graph$ is said to be undirected. 
A path exists between nodes $p,q\in\verts$ if there is a sequence of distinct nodes such that $(v_{0}=p,...,v_{k}=q)$ for $k\in\ZZ_{\geq0}$, $(v_{s-1},v_{s})\in\edges$, and $s\in[k]$.
The graph $\graph$ is said to be connected if there is a path joining any two distinct nodes in $\verts$.
The neighbor set of node $p$ is denoted by $\nbd_p\coloneqq\{q\in\verts\setminus\{p\}:(p,q)\in\edges\}$.
The adjacency matrix of $\graph$ is denoted by $\adj\coloneqq[a_{pq}]\in\RR^{N\times N}$, where $a_{pq}=1$ if and only if $(p,q)\in\edges$, and $a_{pq}=0$ otherwise.
Self-edges are not employed in this work, that is, $a_{pp}\coloneqq 0$ for all $p\in\verts$.
The degree matrix of $\graph$ is denoted by $\Delta\coloneqq\diag{\adj\cdot 1_N}\in\RR^{N\times N}$.
The Laplacian matrix of $\graph$ is denoted by $\lap\coloneqq\Delta-\adj\in\RR^{N\times N}$.
The following result enables the stability analysis provided in Section~\ref{sec: Stability Analysis}.
\begin{lemma} \label{lemma: L and S Identities}
    If $\graph$ is static, undirected, and connected, then there exists an orthonormal basis $\beta\coloneqq\{v_1,v_2,...,v_N\}\subset\RR^N$ for $\Range{\lap}$ such that $v_1 = (\sqrt{N}/N) 1_N$.
    Consider the matrix $\WV\coloneqq [v_2,v_3,...,v_N]\in\RR^{N\times N-1}$ and projection $\sproj\coloneqq I_N - 1_N 1_N^\top / N \in\RR^{N\times N}$.
    Then,\footnote{See~\cite[Appendix A]{Zegers.Phillips2024} for the proof of Lemma~\ref{lemma: L and S Identities}.}
    \begin{align}
        \lap &= \WV\WD\WV^\top, \ \WV^\top\WV = I_{N-1}, \label{eqn: Laplacian Identity} \text{ and }\\
        \sproj &= \WV\WV^\top \label{eqn: S Identity}
    \end{align}
    for some diagonal, positive definite $\WD\in\RR^{N-1\times N-1}$.
    \hfill$\triangle$
\end{lemma}
The matrix $\sproj$ is a projection whose image is the orthogonal complement of the agreement subspace $\Span{1_N}$. 
That is, for any vector $z\in\RR^N$, $z$ may be decomposed into the sum of two orthogonal vectors, i.e., $z = z^\Vert + z^\perp$, where $z^\Vert \coloneqq (1_N 1_N^\top / N)z$ and $z^\perp \coloneqq \sproj z$.

\section{Problem Formulation} \label{sec: Problem Statement}
Consider a MAS of $N\in\ZZ_{\geq 2}$ agents, which are enumerated by the elements of $\verts$.
Let $\theta_p\in\RR$ denote the time kept by agent $p\in\verts$ as defined by a hardware clock with dynamics
\begin{equation} \label{eqn: hardware clock p}
    \dot{\theta}_p \in a_p + \delta_p\ball.
\end{equation}
In~\eqref{eqn: hardware clock p}, $a_p\in\RR_{>0}$ denotes the unknown clock rate of change, which is also commonly referred to as \textit{drift}, and $\delta_p\in [0,a_p)$ denotes a bounding constant for the magnitude of the disturbance affecting the hardware clock of agent $p$.
The hardware clock described by~\eqref{eqn: hardware clock p} resides within the computer of agent $p$ and may exist as a separate integrated circuit.
Let $\vartheta_p\in\RR$ represent the time kept by agent $p$ as defined by a steerable software clock with dynamics
\begin{equation} \label{eqn: software clock p}
    \dot{\vartheta}_p \in a_p + \delta_p\ball + u_p,
\end{equation}
where the control input $u_p\in\RR$ enables time steering.
Since computer programs requiring an awareness of time, such as, task scheduling or timestamping, depend on the timekeeping produced by the computer's hardware clock, any disturbance experienced by the hardware clock is also introduced into the program.
Consequently, the software clock described by~\eqref{eqn: software clock p} inherits the perturbation in~\eqref{eqn: hardware clock p}.
We do not require any stringent smoothness or statistical properties on the perturbations in~\eqref{eqn: hardware clock p} and~\eqref{eqn: software clock p}; we only assume the perturbation affecting agent $p$ is uniformly bounded by a disk of radius $\delta_p$.
Let the constant $a^{\star}\in\RR_{>0}$ represent a desired clock drift which is known to all agents in the MAS.
To facilitate cooperation, the agents may exchange information through a communication network with intermittently available directed components that is supported by a static, undirected, and connected graph $\graph=(\verts,\edges)$.

The objective is to mint a decentralized controller for each agent $p\in\verts$ that achieves the following:
\begin{enumerate}
    \item synchronizes the software clocks within a user-defined tolerance $\nu\in\RR_{>0}$, that is, $\vert\vartheta_p - \vartheta_q\vert\leq \nu$ for all distinct $p,q\in\verts$;
    \item drives all software clock drifts to $a^{\star}$ to ensure synchronization with a desired clock drift;
    \item uses intermittent clock samples from neighboring agents and itself, that is, $\{\vartheta_q\}_{q\in\nbd_p\cup\{p\}}$, which may be procured asynchronously via broadcasts.
\end{enumerate}
Expanding on Item $3)$, the gathering of software clock samples from agents in $\{\vartheta_q\}_{q\in\nbd_p\cup\{p\}}$ may occur at different moments in time and at different rates.
Such a control strategy better uses limited resources relative to continuous communication alternatives, readily integrates in digital hardware, and accommodates the natural asynchronous flow of information within MASs.   

\section{Hybrid System Development} \label{sec: Modeling}
For each $p\in\verts$, let $0<T_1^p\leq T_2^p$ be user-defined constants.
Let $\tau_p\in[0,T_2^p]$ denote the time produced by a software timer of agent $p$ that evolves according to the hybrid system
\begin{equation} \label{eqn: tau_p}
\begin{aligned}
    \dot{\tau}_p &\in -b_p + \delta_p\ball, & \tau_p &\in [0,T_2^p]\\
    \tau_p^+ &\in[T_1^p,T_2^p], & \tau_p &=0
\end{aligned}
\end{equation}
with initial condition satisfying $\tau_p(0,0)\in [T_1^p,T_2^p]$.
Observe, $b_p >\delta_p $ denotes a user-defined timer drift.
Since the perturbed hardware clock described by~\eqref{eqn: hardware clock p} affects all software running in the computer of agent $p$, it may not be possible for $\tau_p$ to exhibit a fixed drift.
Therefore, the differential inclusion in~\eqref{eqn: tau_p} represents a perturbation of the ideal flow equation $\dot{\tau}_p = -b_p$ due to the perturbed flows of $\theta_p$.
The hybrid dynamics in~\eqref{eqn: tau_p} enables the construction of increasing sequences of time, e.g., $\{t_k^p\}_{k=0}^\infty$ given a complete solution $\phi_{\tau_p}$, where the event time $t_k^p$ denotes the $k^\text{th}$ instant $\tau_p=0$.
Consider $b_{p,\max}\coloneqq b_p + \delta_p \in\RR_{>0}$ and $b_{p,\min}\coloneqq b_p - \delta_p \in\RR_{>0}$.
One can then show the event times generated by~\eqref{eqn: tau_p} satisfy the following inequalities: for all $k\in\ZZ_{\geq 0}$,
\begin{equation} \label{eqn: timer_p bounds}
    \frac{T_1^p}{b_{p,\max}} \leq t_{k+1}^p - t_k^p \leq \frac{T_2^p}{b_{p,\min}}.
\end{equation}
The timers $\{\tau_p\}_{p\in\verts}$ can be treated as independent autonomous systems and will trigger specific actions for their corresponding owner, that is, $\tau_p$ belongs to agent $p$ for each $p\in\verts$.
In particular, $\tau_p$ will be used to dictate when agent $p$ broadcasts or pushes its software clock value from itself to its neighbors.
To simplify the preliminary development, we suppose broadcasts occur in zero continuous-time without dropouts---when agent $p$ broadcasts information, all neighbors of agent $p$ instantaneously and simultaneously receive the broadcast information.
Since radio and optical communications utilize electromagnetic waves traveling at the speed of light, such an assumption is reasonable when agents communicate over short distances with unobstructed lines of sight.\footnote{The speed of light is approximately $3\times 10^8$ m/s.
If the sending of information from an emitter to a receiver within $1\times 10^{-4}$ s defines instantaneous communication, then a quick calculation reveals that a receiver and emitter can be separated by at most $30$ km or about $18$ miles to experience instantaneous communication.
These distances may need to be slightly reduced to lessen the probability of packet dropouts and account for the encoding and decoding of information.}
When inter-agent distances become large, e.g., when communication takes place beyond the local horizon or in cislunar space, communication delays, packet dropouts, and encoding-decoding time must be considered; we reserve such challenges for future work.

For every $p\in\verts$, let $\hat{\vartheta}_p\in\RR$ be an auxiliary variable that evolves according to
\begin{equation} \label{eqn: varthetaHat_p dynamics}
\begin{aligned}
	\dot{\hat{\vartheta}}_p &= a^{\star}, & \tau_p &\in[0,T_2^p]\\
	\hat{\vartheta}_p^+ &= \vartheta_p, & \tau_p &= 0.
\end{aligned}
\end{equation}
The hybrid system in~\eqref{eqn: varthetaHat_p dynamics} states that whenever a jump is caused by $\tau_p=0$, the variable $\hat{\vartheta}_p$ is reset to the instantaneous value of $\vartheta_p$, and this updated value for $\hat{\vartheta}_p$ will serve as the initial condition for the initial value problem (IVP) defined by the differential equation in~\eqref{eqn: varthetaHat_p dynamics} during flows.
Thus, \eqref{eqn: varthetaHat_p dynamics} describes an IVP that is reset according to the event times generated by $\tau_p$, which can occur intermittently or periodically depending on the selection of $T_1^p$ and $T_2^p$.
If the parameters $T_1^p$ and $T_2^p$ are selected such that $0<T_1^p<T_2^p$ holds, then the times when $\vartheta_p$ is sampled and broadcast may be intermittent.
Nevertheless, periodic sampling and broadcasting can be created by selecting $0<T_1^p=T_2^p$.

For every agent $p\in\verts$, the parameter $a_p$ is unknown.
Yet, this parameter can be reconstructed using
\begin{align}
    \dot{\hat{a}}_p &= k_a(\theta_p - \hat{\theta}_p), \label{eqn: ahatDot_p} \\
    \dot{\hat{\theta}}_p &= \hat{a}_p + k_{\theta}(\theta_p - \hat{\theta}_p), \label{eqn: thetahatDot_p}
\end{align}
where $k_a\in\RR_{>0}$ and $k_{\theta}\in\RR_{>0}$ are user-defined parameters, $\hat{a}_p\in\RR$ denotes agent $p$'s estimate of $a_p$, and $\hat{\theta}_p\in\RR$ denotes agent $p$'s estimate of $\theta_p$.
Although the values of the hardware clock of agent $p$ (i.e., $\theta_p$) are measurable, the estimate $\hat{\theta}_p$ is used to create a feedback signal that enables the reconstruction of $a_p$.
Given a user-defined parameter $k_u\in\RR_{>0}$, the controller of agent $p\in\verts$ is designed as
\begin{equation} \label{eqn: agent controller}
    u_p \coloneqq a^{\star} - \hat{a}_p + k_u\sum_{q\in\nbd_p}\big(\hat{\vartheta}_q - \hat{\vartheta}_p\big)\in\RR.
\end{equation}
The controller in~\eqref{eqn: agent controller} is distributed as it only uses information from neighboring agents and the implementing agent itself.
In addition, the controller is amenable to decentralized implementation given the systems in~\eqref{eqn: tau_p}-\eqref{eqn: thetahatDot_p} and the communication assumption made above~\eqref{eqn: varthetaHat_p dynamics}.
Under this construction, the variables $\hat{\vartheta}_q$ (one for each agent $q\in\nbd_p\cup\{p\}$) are instantaneously and simultaneously updated if and only if an event is triggered by $\tau_q = 0$---it is the fact that agents push information, rather than pull information as in~\cite{Guarro.Sanfelice2023}, that facilitates decentralization.
It is also worth noting that the values of $\hat{\vartheta}_q$ and $\hat{\vartheta}_p$ may be reset asynchronously since the timers $\tau_q$ and $\tau_p$ are independent. 

To facilitate the derivation of the closed-loop hybrid system used to model the behavior of the ensemble, let
\begin{align}
    \tilde{\vartheta}_p &\coloneqq \vartheta_p - \hat{\vartheta}_p \in\RR, \label{eqn: varthetaTilde_p}\\
    \tilde{a}_p &\coloneqq a_p - \hat{a}_p\in\RR, \label{eqn: aTilde_p} \\
    \tilde{\theta}_p &\coloneqq \theta_p - \hat{\theta}_p\in\RR \label{eqn: thetaTilde_p}.
\end{align}
We provide the following notation to aid the writing of concise expressions.
Let
\begin{equation*}
\begin{aligned}
    \vartheta &\coloneqq (\vartheta_p)_{p\in\verts}\in\RR^N, \quad \theta \coloneqq (\theta_p)_{p\in\verts}\in\RR^N, \\
    \hat{\vartheta} &\coloneqq (\hat{\vartheta}_p)_{p\in\verts}\in\RR^N, \quad \hat{\theta} \coloneqq (\hat{\theta}_p)_{p\in\verts}\in\RR^N,\\
    \tilde{\vartheta} &\coloneqq (\tilde{\vartheta}_p)_{p\in\verts}\in\RR^N, \quad \tilde{\theta} \coloneqq (\tilde{\theta}_p)_{p\in\verts}\in\RR^N,\\
    a &\coloneqq (a_p)_{p\in\verts}\in\RR^N, \quad \tau \coloneqq (\tau_p)_{p\in\verts}\in\RR^N.\\
    \hat{a} &\coloneqq (\hat{a}_p)_{p\in\verts}\in\RR^N, \\
    \tilde{a} &\coloneqq (\tilde{a}_p)_{p\in\verts}\in\RR^N, \\
\end{aligned}
\end{equation*}
Given the development above, we now derive equations and inclusions leading to the flow map of the closed-loop ensemble hybrid system.
The substitution of~\eqref{eqn: agent controller}--\eqref{eqn: aTilde_p} into~\eqref{eqn: software clock p} yields
\begin{equation} \label{eqn: varthetaDot_p}
    \dot{\vartheta}_p \!\in a^{\star} + \tilde{a}_p + k_u\sum_{q\in\nbd_p} \big(\tilde{\vartheta}_p - \tilde{\vartheta}_q \big) + k_u\sum_{q\in\nbd_p} \big(\vartheta_q - \vartheta_p\big) + \delta_p\ball.
\end{equation}
The substitution of the flow equation in~\eqref{eqn: varthetaHat_p dynamics} and~\eqref{eqn: varthetaDot_p} into the set-valued derivative of~\eqref{eqn: varthetaTilde_p} leads to the differential inclusion
\begin{equation} \label{eqn: varthetaTildeDot_p}
    \dot{\tilde{\vartheta}}_p \!\in \tilde{a}_p + k_u\sum_{q\in\nbd_p} \big(\tilde{\vartheta}_p - \tilde{\vartheta}_q \big) + k_u\sum_{q\in\nbd_p} \big(\vartheta_q - \vartheta_p\big) + \delta_p\ball.
\end{equation}
The substitution of~\eqref{eqn: ahatDot_p} and~\eqref{eqn: thetaTilde_p} into the time derivative of~\eqref{eqn: aTilde_p} yields
\begin{equation} \label{eqn: aTildeDot_p}
    \dot{\tilde{a}}_p = -k_a\tilde{\theta}_p. 
\end{equation}
The substitution of~\eqref{eqn: hardware clock p}, \eqref{eqn: thetahatDot_p}, \eqref{eqn: aTilde_p}, and~\eqref{eqn: thetaTilde_p} into the set-valued derivative of~\eqref{eqn: thetaTilde_p} leads to the inclusion
\begin{equation} \label{eqn: thetaTildeDot_p}
    \dot{\tilde{\theta}}_p \in \tilde{a}_p -k_{\theta}\tilde{\theta}_p + \delta_p\ball. 
\end{equation}
Substituting~\eqref{eqn: varthetaDot_p} into the set-valued derivative of $\vartheta$ for each $p\in\verts$ while using the definitions of $\tilde{a}$, $\vartheta$, and $\tilde{\vartheta}$ yields
\begin{equation} \label{eqn: varthetaDot}
    \dot{\vartheta} \in a^{\star}1_N + \tilde{a} + k_u\lap\tilde{\vartheta} - k_u\lap\vartheta + \{(d_p)_{p\in\verts}\colon \forall_{p\in\verts} \ d_p\in\delta_p\ball  \}.
\end{equation}
Substituting the flow equation in~\eqref{eqn: varthetaHat_p dynamics} into the time derivative of $\hat{\vartheta}$ for every $p\in\verts$ results in
\begin{equation} \label{eqn: varthetaHatDot}
    \dot{\hat{\vartheta}} = a^{\star}1_N.
\end{equation}
Substituting~\eqref{eqn: aTildeDot_p} into the time derivative of $\tilde{a}$ for each $p\in\verts$ while using the definition of $\tilde{\theta}$ yields
\begin{equation} \label{eqn: aTildeDot}
    \dot{\tilde{a}} = -k_a\tilde{\theta}.
\end{equation}
Substituting~\eqref{eqn: thetaTildeDot_p} into the set-valued derivative of $\tilde{\theta}$ for each $p\in\verts$ while using the definitions of $\tilde{a}$ and $\tilde{\theta}$ yields
\begin{equation} \label{eqn: thetaTildeDot}
    \dot{\tilde{\theta}} \in \tilde{a} -k_{\theta}\tilde{\theta} + \{(d_p)_{p\in\verts}\colon \forall_{p\in\verts} \ d_p\in\delta_p\ball  \}.
\end{equation}

To measure the degree of synchronization between the times in $\vartheta$, we will require a disagreement metric.
One potential metric can be constructed from a projection.
In $\RR^N$, the agreement subspace is denoted by $\agreement\coloneqq \{\vartheta\in\RR^N\colon \forall_{p\in\verts} \ \vartheta_p = \vartheta_q \}$.
Since $\agreement$ is a subspace, every configuration $\vartheta$ can be decomposed into orthogonal components, namely, $\vartheta^\Vert = (1_N 1_N^\top/N)\vartheta\in\RR^N$ and $\vartheta^\perp = \sproj\vartheta\in\RR^N$, where $\vartheta = \vartheta^\Vert + \vartheta^\perp$.
Note, $( \vartheta^\Vert)^\top \vartheta^\perp = 0$ follows by direct computation.
Moreover, $\vartheta^\Vert$ and $\vartheta^\perp$ represent the components of $\vartheta$ in agreement and disagreement, respectively. 
Given the objective, Item $1)$ can be achieved by driving the disagreement $\Vert\vartheta^\perp\Vert$ to zero.
However, one can also formulate an alternative disagreement metric which is more convenient than $\Vert\vartheta^\perp\Vert$.
The following result is a modification of~\cite[Lemma 2]{Zegers.Phillips2024}.
\begin{lemma} \label{lemma: Equivalent Consensus Metric}
    For any $z\in\RR^N$, $\Vert z^\perp \Vert = \Vert \WV^\top z\Vert$.\footnote{The definition of $\WV$ is provided in Lemma~\ref{lemma: L and S Identities}.} 
    \hfill$\triangle$
\end{lemma}
\begin{proof}
    Given Lemma~\ref{lemma: L and S Identities}, $\sproj = \WV\WV^\top$.
   Furthermore, $\sproj = \sproj^\top$ by construction, and $\sproj^2 = \sproj$ since $\sproj$ is a projection.
    Therefore,
    \begin{equation*}
        \Vert z^\perp \Vert^2 = \Vert \sproj z \Vert^2 = z^\top \sproj^\top \sproj z = z^\top \sproj z = z^\top \WV\WV^\top z = \Vert \WV^\top z \Vert^2,
    \end{equation*}
    and the desired result follows.
\end{proof}
\noindent In light of Lemma~\ref{lemma: Equivalent Consensus Metric}, let 
\begin{equation} \label{eqn: eta}
    \eta\coloneqq\WV^\top \vartheta\in\RR^{N-1}.
\end{equation} 
Since the distance between $\vartheta$ and $\agreement$ is quantified by $\Vert \vartheta^\perp \Vert$ and $\Vert \vartheta^\perp \Vert = \Vert \eta \Vert$ by Lemma~\ref{lemma: Equivalent Consensus Metric} and~\eqref{eqn: eta}, it follows that $\Vert \eta \Vert$ is an alternative metric for disagreement.
Therefore, $\vartheta_p=\vartheta_q$ for all distinct $p,q\in\verts$ if and only if $\Vert \eta \Vert = 0$.
Note, $\Vert \eta \Vert$ captures the disagreement between between all software-defined times.
The closed-loop dynamics of $\eta$ can be obtained by substituting~\eqref{eqn: Laplacian Identity} and~\eqref{eqn: varthetaDot} into the set-valued derivative of~\eqref{eqn: eta}, where
\begin{equation} \label{eqn: etaDot}
    \dot{\eta} \in \WV^\top\tilde{a} + k_u\WD\WV^\top \tilde{\vartheta} - k_u\WD\eta + \big\{\WV^\top(d_p)_{p\in\verts}\colon \forall_{p\in\verts} \ d_p\in \delta_p\ball \big\}.
\end{equation}
Substituting~\eqref{eqn: Laplacian Identity}, \eqref{eqn: varthetaDot}, \eqref{eqn: varthetaHatDot}, and~\eqref{eqn: eta} into the set-valued derivative of $\tilde{\vartheta}$ leads to the differential inclusion
\begin{equation} \label{eqn: varthetaTildeDot}
    \dot{\tilde{\vartheta}} \in \tilde{a} + k_u\lap\tilde{\vartheta} - k_u\WV\WD\eta + \{(d_p)_{p\in\verts}\colon \forall_{p\in\verts} \ d_p\in\delta_p\ball  \}.
\end{equation}
With these objects in place, we can now define a hybrid system for the ensemble.

Let $\hyb$ be a closed-loop hybrid system for the MAS, where $\xi\coloneqq (\eta,\tilde{\vartheta},\tilde{a},\tilde{\theta},\tau)\in\statespc$ and $\statespc\coloneqq \RR^{N-1}\times\RR^N\times\RR^N\times\RR^N\times\RR^N$ denote the state vector and state space, respectively.
The flow set of the hybrid system $\hyb$ is
\begin{equation*}
    C \coloneqq \bigcap_{p\in\verts} \{\xi\in\statespc\colon \tau_p\in[0,T_2^p]\}.
\end{equation*}
Let $z\coloneqq (\eta,\tilde{\vartheta,}\tilde{a},\tilde{\theta})$ be an auxiliary variable, which enables the writing of $\xi=(z,\tau)$.
Moreover, consider the perturbation set 
\begin{equation*}
    P_z\coloneqq \{(\WV^\top(d_p)_{p\in\verts},(d_p)_{p\in\verts},0_N,(d_p)_{p\in\verts})\colon \forall_{p\in\verts} \ d_p\in \delta_p\ball \}.
\end{equation*}
The substitution of~\eqref{eqn: aTildeDot}, \eqref{eqn: thetaTildeDot}, \eqref{eqn: etaDot}, and~\eqref{eqn: varthetaTildeDot} into the generalized time derivative of $z$ yields $\dot{z} \in \fmat z + P_z$, where
\begin{equation} \label{eqn: F map}
\begin{aligned}
    \fmat \coloneqq 
    \left[
    \begin{array}{cccc}
        -k_u\WD & k_u\WD\WV^\top & \WV^\top & 0_{N-1\times N} \\
        -k_u\WV\WD & k_u\lap & I_N & 0_{N\times N} \\
        0_{N\times N-1} & 0_{N\times N} & 0_{N\times N} &  -k_a I_N \\
        0_{N\times N-1} & 0_{N\times N} & I_N &  -k_{\theta} I_N \\
    \end{array}
    \right].
\end{aligned}
\end{equation}
Let $P_{\tau}\coloneqq \{(d_p)_{p\in\verts}\colon \forall_{p\in\verts} \ d_p\in\delta_p\ball \}$ denote the perturbation set corresponding to $\tau$.
The flows of $\hyb$ are governed by the set-valued map $F\colon\statespc\rightrightarrows\statespc$, where $\dot{\xi}\in F(\xi)$ and 
\begin{equation} \label{eqn: flow map}
    F(\xi)\coloneqq (\fmat z + P_z)\times (-(b_p)_{p\in\verts} + P_{\tau}).
\end{equation}
Note, the derivation of $F$ follows from the definitions of $\xi$ and $z$, the flow inclusion in~\eqref{eqn: tau_p} for every $p\in\verts$ and~\eqref{eqn: F map}.  
The jump set of the hybrid system $\hyb$ is
\begin{equation*}
    D \coloneqq \bigcup_{p\in\verts} \{\xi\in\statespc\colon \tau_p=0\}.
\end{equation*}
For each agent $p\in\verts$, let $D_p \coloneqq \{\xi\in\statespc\colon \tau_p=0\}$.
The jumps of $\hyb$ are governed by the set-valued map $G\colon\statespc\rightrightarrows\statespc$, where $\xi^+ \in G(\xi)$ and
\begin{equation} \label{eqn: jump map}
\begin{aligned}
    G(\xi) &\coloneqq \big\{G_p(\xi)\colon\xi\in D_p\text{ for some }p\in\verts\big\}, \\
    G_p(\xi) &\coloneqq 
    \{\eta\}\times\{(\tilde{\vartheta}_q^+)_{q\in\verts}\}\times\{\tilde{a}\}\times\{\tilde{\theta}\}\times\TT_p(\tau), \\
    \TT_p(\tau) &\coloneqq \{\tau_1\}\!\times\!...\!\times\!\{\tau_{p-1}\}\!\times\![T_1^p,T_2^p]\!\times\!\{\tau_{p+1}\}\!\times\! ...\!\times\!\{\tau_N\}, \\
    \tilde{\vartheta}_q^+ &= \left\{
        \begin{aligned}
            &\tilde{\vartheta}_q, & q &\neq p \\
            &0, & q &= p.
        \end{aligned}
    \right. 
\end{aligned}
\end{equation}
The jump map in~\eqref{eqn: jump map} is derived by employing the following observations.
Utilizing~\eqref{eqn: software clock p} for each $p\in\verts$, the definition of $\vartheta$, and~\eqref{eqn: eta}, one can show $\eta$ evolves only in continuous-time.
Thus, $\eta^+=\eta$ under any jump.
Using similar arguments, one can also show that $\tilde{a}$ and $\tilde{\theta}$ evolve only in continuous-time.
Therefore, $\tilde{a}^+ = \tilde{a}$ and $\tilde{\theta}^+ = \tilde{\theta}$ under any jump.
However, the jump inclusion in~\eqref{eqn: tau_p} implies $\tau_p^+\in [T_1^p,T_2^p]$ in response to a jump triggered by $\tau_p = 0$; otherwise, $\tau_p^+ = \tau_p$ in response to a jump not triggered by $\tau_p = 0$.
The flow inclusion in~\eqref{eqn: software clock p}, the jump equation in~\eqref{eqn: varthetaHat_p dynamics}, and the definition in~\eqref{eqn: varthetaTilde_p} imply $\tilde{\vartheta}_p^+=0$ whenever a jump occurs in response to $\tau_p = 0$, and $\tilde{\vartheta}_p^+=\tilde{\vartheta}_p$ otherwise.

The solutions of the hybrid system $\hyb$ with data $(C,F,D,G)$ describe the behavior of the ensemble.
Consequently, the MAS accomplishes synchronization in $\{\vartheta_p\}_{p\in\verts}$ if the set
\begin{equation} \label{eqn: attractor}
    \attr \coloneqq \big\{\xi\in C\colon \Vert z \Vert = 0 \big\}
\end{equation}
is GES---a sufficient condition.
Lamentably, the perturbations influencing the hardware clocks, software clocks, and software timers challenge the GES of $\attr$ for $\hyb$.
Yet, depending on the application, there may be an acceptable deviation from perfect synchronization, i.e., $\eta= 0_{N-1}$.
This motivates the following definition.
\begin{definition} \label{def: nu-approximate synchronization}
    Let $\nu\in\RR_{>0}$ be a synchronization tolerance.
    The hybrid system $\hyb$ specified by the data $(C,F,D,G)$ is said to achieve $\nu$-approximate synchronization in the software-defined times $\{\vartheta_p\}_{p\in\verts}$ if, for every \update{maximal} solution $\phi$ of $\hyb$, there exists a $T\in\RR_{\geq 0}$ such that $\vert\phi_{\vartheta_p}(t,j) - \phi_{\vartheta_q}(t,j)\vert\leq\nu$ for all $(p,q)\in\edges$ and $t+j\geq T$ with $(t,j)\in\dom\phi$.
    \hfill$\triangle$
\end{definition}
Given Definition~\ref{def: nu-approximate synchronization}, it will be suitable to employ the uniform norm, namely, 
\begin{equation}
    \Vert \vartheta\Vert_{\infty} \coloneqq \max\{\vert \vartheta_p -\vartheta_q\vert\colon (p,q)\in\edges \},
\end{equation}
where we remind the reader that $\edges$ denotes the edge set of the coupling graph $\graph$.
Utilizing Lemma~\ref{lemma: Equivalent Consensus Metric}, \eqref{eqn: eta}, and~\cite[Equation 21]{Zegers.Guralnik.ea2024}, one can obtain
\begin{equation} \label{eqn: norm equivalence}
    \frac{1}{\sqrt{N}}\Vert \eta\Vert \leq \Vert \vartheta \Vert_{\infty} \leq \sqrt{2}\Vert \eta \Vert.
\end{equation}
Note, if $\Vert z\Vert\leq \nu/\sqrt{2}$, then $\Vert\vartheta\Vert_{\infty} \leq \nu$ because the definition of $z$ implies $\Vert\eta\Vert\leq \Vert z\Vert$ and~\eqref{eqn: norm equivalence} implies $\Vert\vartheta\Vert_{\infty}\leq \sqrt{2}\Vert\eta\Vert$.
Let $\tspc\coloneq [0,T_2^1]\times[0,T_2^2]\times ...\times[0,T_2^N]$ be an auxiliary space.
Given~\eqref{eqn: attractor}, $\xi^\prime\in\attr$ if and only if $\xi^\prime = (\mathbf{0},\tau^\prime)$ and $\tau^\prime\in\tspc$.
As a result, for any $\xi\in C\cup D$,
\begin{equation} \label{eqn: distance to desired set}
    \vert \xi \vert_{\attr} = \inf\{\Vert \xi - \xi^\prime\Vert\colon\xi^\prime\in\attr\} = \Vert z \Vert.
\end{equation}
In addition, the hybrid system $\hyb$ accomplishes $\nu$-approximate synchronization in $\{\vartheta_p\}_{p\in\verts}$ if the set $\{\xi\in\statespc\colon \vert\xi\vert_{\attr}\leq \nu/\sqrt{2}\}$ is attractive for $\hyb$---this observation motivates the following definition.
\begin{definition} \label{def: GPES}
    A closed set $\attr\subset C\cup D$ is said to be globally practically exponentially stable (GPES) for the hybrid system $\hyb$ with data $(C,F,D,G)$ if, for each \update{maximal} solution $\phi$ of $\hyb$, \update{$\phi$ is complete and} there exist constants $\alpha,\kappa_1,\kappa_2\in\RR_{>0}$ such that
    \begin{equation*}
        \vert\phi(t,j)\vert_{\attr} \leq \kappa_1\exp(-\alpha (t+j)) + \kappa_2
    \end{equation*}
    for all $(t,j)\in\dom\phi$.\footnote{Solutions of $\hyb$ are not defined outside of $C\cup D\cup G(D)$.
    Thus, the global qualifier in GPES refers to the entire space $C\cup D\cup G(D)$ as the complement is immaterial.}
    \hfill$\triangle$
\end{definition}
\noindent By Definitions~\eqref{def: nu-approximate synchronization} and~\eqref{def: GPES}, the $\nu$-approximate synchronization problem can be recast as a set stabilization problem for hybrid systems provided $\kappa_2 < \nu/\sqrt{2}$. 

Under the construction of $\hyb$, the flow set $C$ and jump set $D$ are closed.
The flow map $F$ is outer semi-continuous, locally bounded, and convex-valued.
The jump map $G$ is outer semi-continuous and locally bounded.
As a result, $\hyb$ satisfies the hybrid basic conditions~\cite[Assumption 6.5]{Goebel.Sanfelice.ea2012}, and~\cite[Theorem 6.8]{Goebel.Sanfelice.ea2012} implies $\hyb$ is nominally well-posed.

\section{Stability Analysis} \label{sec: Stability Analysis}
We begin this section by demonstrating that every maximal solution $\phi$ of $\hyb$ is complete, which ensures the existence of $\vert\phi(t,j)\vert_{\attr}$ for arbitrarily large $t+j$ such that $(t,j)\in\dom\phi$.
Note, it will be useful to define the following constants:
\begin{equation*}
\begin{aligned}
    T_{\min} &\coloneqq \min \{T_1^p\colon p\in\verts\}, \ \ T_{\max} \coloneqq \max \{T_2^p\colon p\in\verts\}, \\
    b_{\min} &\coloneqq \min \{b_{p,\min}\colon p\in\verts\}, \ \ b_{\max} \coloneqq \max \{b_{p,\max}\colon p\in\verts\}.
\end{aligned}
\end{equation*}
\begin{lemma} \label{lemma: Completeness and (t,j) Bounds}
    For every maximal solution $\phi$ of the hybrid system $\hyb$ with data $(C,f,D,G)$, the following items hold:
    \begin{enumerate}
        \item $\phi$ is complete and non-Zeno.
        \item Along the solution $\phi$, the hybrid dynamics of $\tau$ imply
        \begin{equation} \label{eqn: (t,j) inequality}
           \left(\frac{j}{N} - 1\right)\frac{T_{\min}}{b_{\max}} \leq t \leq \frac{j T_{\max}}{N b_{\min}}.
        \end{equation}
        for all $(t,j)\in\dom\phi$.
        \hfill$\triangle$
    \end{enumerate}
\end{lemma}
\begin{proof}
    Item 1) Let $\phi$ be a maximal solution of $\hyb$ and recall the data $(C,F,D,G)$.
    If $\phi(0,0)\in D$, then $\phi$ experiences at least one jump and lands in $C$, i.e., $\phi(0,k)\in C$ for $k\in\ZZ_{\geq 1}$, which implies $\dom\phi$ contains at least two distinct points.
    If $\phi(0,0)\in C\setminus D$ and $\mathcal{U}$ is a neighborhood of $\phi(0,0)$, then for every $\xi^{\prime}\in C\cap\mathcal{U}$ one can demonstrate that $F(\xi^{\prime})\cap T_C(\xi^{\prime})\neq\varnothing$---use the definition of the tangent (contingent) cone $T_C(\xi)$,
    \begin{equation*}
         \frac{\xi_n^{\prime} - \xi^{\prime}}{\lambda_n} = \frac{(\xi_z^{\prime}+\frac{1}{n}\mathsf{F}\xi_z,\xi_{\tau_1}^{\prime}-\frac{c_1}{n},...,\xi_{\tau_N}^{\prime}-\frac{c_N}{n})-(\xi_z^{\prime},\xi_{\tau}^{\prime})}{\frac{1}{n}},
    \end{equation*}
    select $c_p\in -b_p + \delta_p\ball$ for all $p\in\verts$, and select $n\in\ZZ_{\geq N}$ with $N\in\ZZ_{\geq 1}$ sufficiently large so that $\xi_n^{\prime}\in\mathcal{U}$ for every $n\geq N$.
    Thus, there exists a nontrivial solution for each $\phi(0,0)\in C\cup D$.
    Given the design of $\hyb$, we can see that $F$ in~\eqref{eqn: flow map} has linear growth on $C$ (see~\cite[Definition A.28]{Sanfelice2021b}) and $G(D)\subset C\cup D$.
    Hence, \cite[Proposition 2.34]{Sanfelice2021b} implies every maximal solution of $\hyb$ is complete.
    Item 2) follows from a similar proof to that for~\cite[Lemma 3]{Zegers.Phillips2024}, which implies complete solutions are non-Zeno.
\end{proof}
\vspace{-3pt}

In anticipation of the main result, we introduce the following items.
Let $T_2 \coloneqq (T_2^p)_{p\in\verts}\in\RR^N$ and $\sigma\in\RR_{>0}$ be a constant.
Furthermore, let $P_1\in\RR^{N-1\times N-1}$ be a symmetric, positive definite matrix, $P_{2,k}\in\RR_{>0}$ be a constant for each $k\in\verts$, and $P_3\in\RR^{2N\times 2N}$ a symmetric, positive definite matrix.
Further, let
\begin{equation} \label{eqn: P2(tau)}
\begin{aligned}
    P_2(\tau) &\coloneqq \diag{P_{2,1}\exp(\sigma\tau_1),P_{2,2}\exp(\sigma\tau_2),..., \\
    &\phantom{=} \ \ P_{2,N}\exp(\sigma\tau_N)}\in\RR^{N\times N}.
\end{aligned}
\end{equation}
These objects can be used to define additional items that will assist a Lyapunov-based analysis, namely,
\begin{equation} \label{eqn: theorem 1 items}
\begin{aligned}
    P(\tau) & \coloneqq \diag{P_1,P_2(\tau),P_3}, \\
    Q(\tau) &\coloneqq -\sigma b_{\min} \diag{0_{N-1\times N-1}, P_2(\tau), 0_{2N\times 2N}}, \\
    \mmat(\tau) &\coloneqq \fmat^\top P(\tau) + P(\tau)\fmat + Q(\tau), \\
    \mu &\coloneqq -\sup\{\lambda_{\max}(\mmat(\tau))\colon \tau\in\tspc\}, \\
    \alpha_1 &\coloneqq \lambda_{\min}(P(0_N)), \ \alpha_2\coloneqq \lambda_{\max}(P(T_2)).
\end{aligned}
\end{equation}
By definition, $\mmat(\tau)$ is symmetric for all $\tau\in\tspc$, which implies the eigenvalues of $\mmat(\tau)$ are real.
If $\mmat(\tau)$ is negative definite for all $\tau\in\tspc$, then the constant $\mu$ is positive.
Also, $\alpha_1\in\RR_{>0}$ and $\alpha_2\in\RR_{>0}$ since $P(\tau)$ is symmetric and positive definite.

\begin{theorem} \label{thm: A is GPES} 
If there exist timer parameters $0<T_1^p\leq T_2^p$ for every $p\in\verts$, a constant $\sigma > 0$, controller gain $k_u > 0$, estimator gains $k_a, k_{\theta} > 0$, and symmetric, positive definite matrices $P_1$, $P_2(\tau)$, and $P_3$ that ensure the symmetric matrix $\mmat(\tau)$ is negative definite for all $\tau\in\tspc$, then the set $\attr$ in~\eqref{eqn: attractor} is GPES for the hybrid system $\hyb$ with data $(C,f,D,G)$.
In particular, for every maximal solution $\phi$ of $\hyb$, it follows that
\begin{equation} \label{eqn: GPES bound}
    \vert \phi(t,j) \vert_{\attr} \leq \kappa_1\exp(-\alpha(t+j))\vert \phi(0,0)\vert_{\attr} + \kappa_2
\end{equation}
for all $(t,j)\in\dom\phi$, where, for some constant $\varepsilon\in(0,1)$, 
\begin{equation*} 
\begin{aligned}
    \kappa &\in (0,\mu/\Vert P(T_2)\Vert), \quad \bar{\mu}\coloneqq \left(\mu-\kappa\Vert P(T_2)\Vert\right)/\alpha_2 \in\RR_{>0}\\
    \kappa_1 &\coloneqq \sqrt{\frac{\alpha_2}{\alpha_1}\exp\big(\bar{\mu}(1-\varepsilon)T_{\min}\big)}\vert\phi(0,0)\vert_{\attr} \in\RR_{>0}, \\
    \kappa_2 &\coloneqq \sqrt{\frac{\Vert P(T_2)\Vert }{\alpha_1 \bar{\mu}\kappa}}\delta_{\max} \in\RR_{>0}, \ \delta_{\max}\coloneqq \sup \{\Vert d \Vert\colon d\in P_z\}, \\
    \alpha &\coloneqq \frac{1}{2}\min\left\{ \bar{\mu}\varepsilon,\frac{\bar{\mu}(1-\varepsilon)T_{\min}}{N}\right\}\in\RR_{>0}.
\end{aligned}
\end{equation*}
If $\nu/\sqrt{2}\in (\kappa_2,\kappa_1\vert\phi(0,0)\vert_{\attr} + \kappa_2)$, then the hybrid system $\hyb$ achieves $\nu$-approximate synchronization in $\{\vartheta_p\}_{p\in\verts}$ with 
\begin{equation} \label{eqn: nu time}
    T = \frac{1}{\alpha}\ln\left(\frac{\sqrt{2}\kappa_1\vert \phi(0,0)\vert_{\attr}}{\nu - \sqrt{2}\kappa_2}\right).
\end{equation}
Moreover, the control trajectories $\{\phi_{u_p}(t,j)\}_{p\in\verts}$ are bounded for all $(t,j)\in\dom\phi$.
\hfill$\triangle$
\end{theorem}
\begin{proof}
First, observe $\statespc$ is an open set, such that $\attr\subset\statespc$ and $C\cup D\cup G(D)\subset\statespc$.
Consider the Lyapunov function candidate
\begin{equation} \label{eqn: V}
    V\colon\statespc\to\RR_{\geq 0}\colon \xi \mapsto
    z^\top P(\tau) z.
\end{equation}
Utilizing $\vert\xi\vert_{\attr}=\Vert z\Vert$ by way of~\eqref{eqn: distance to desired set} and the definitions of $\alpha_1$ and $\alpha_2$ in~\eqref{eqn: theorem 1 items}, the Lyapunov function candidate in~\eqref{eqn: V} can be bounded as
\begin{equation} \label{eqn: V Bounds}
    \alpha_1\vert\xi\vert_{\attr}^2 \leq V(\xi)\leq \alpha_2\vert\xi\vert_{\attr}^2.
\end{equation}
When $\xi\in C$, $\xi$ evolves according to $\dot{\xi}\in F(\xi)$.
Recall, the map $F$ is defined in~\eqref{eqn: flow map}.
Since the temporal evolution of $\xi$ is described by a differential inclusion, we require an appropriate generalization for the time derivative of $V(\xi)$.
While several extensions of the standard time derivative exist, we will utilize the following notion:
\begin{equation} \label{eqn: generalized time derivative}
    \dot{V}(\xi) \coloneqq \max_{\mathfrak{f}\in F(\xi)}\nabla V(\xi)^\top \mathfrak{f},
\end{equation}
which is well-defined since~\eqref{eqn: V} is continuously differentiable in $\xi$.
For each $\xi\in\statespc$, the set $F(\xi)$ is compact by construction.
Also, the linear function $h(\_;\xi)\colon F(\xi)\to\RR\colon \mathfrak{f}\mapsto \nabla V(\xi)^\top\mathfrak{f}$ is continuous.
Therefore, $h(\_;\xi)$ attains a maximum over $F(\xi)$ for all $\xi\in\statespc$.
It then follows that, for each $\xi\in C$ there exists a $d_z\in P_z$ and a $d_{\tau}\coloneqq (d_{\tau,p})_{p\in\verts}\in P_{\tau}$, such that
\begin{equation} \label{eqn: VDot 1} 
\begin{aligned}
    \dot{V}(\xi) &= \max_{\mathfrak{f}\in F(\xi)}\nabla V(\xi)^\top\mathfrak{f}  \\
    &= 2z^\top P(\tau)(\fmat z + d_z) \\ 
    &\phantom{=} \ + \sum_{p\in\verts} \sigma\vartheta_p P_{2,p}\exp(\sigma\tau_p)\vartheta_p (-b_p + d_{\tau,p}) \\
    &\leq 2z^\top P(\tau)(\fmat z + d_z) - \sigma b_{\min} \sum_{p\in\verts}\vartheta_p P_{2,p}\exp(\sigma\tau_p)\vartheta_p \\
    &= z^\top \big( \fmat^\top P(\tau) + P(\tau)\fmat + Q(\tau) \big) z + 2z^\top P(\tau)d_z \\
    &= z^\top \mmat(\tau) z + 2z^\top P(\tau)d_z.
\end{aligned}
\end{equation}
Since $\mmat(\tau)$ is negative definite for all $\tau\in\tspc$ by the hypothesis, the constant $\mu$ defined in~\eqref{eqn: theorem 1 items} is positive.
Therefore, the definition of $\mu$, $\vert\xi\vert_{\attr}=\Vert z\Vert$, $V(\xi)\leq \alpha_2\vert\xi\vert_{\attr}^2$ from~\eqref{eqn: V Bounds}, and the final inequality in~\eqref{eqn: VDot 1} imply
\begin{equation} \label{eqn: VDot 2}
\begin{aligned}
    \dot{V}(\xi) &\leq -\mu\Vert z\Vert^2 + 2z^\top P(\tau)d_z \\ 
    &\leq -\mu\vert\xi\vert_{\attr}^2 + 2\sup_{\tau\in\tspc}\Vert P(\tau)\Vert \cdot\Vert z \Vert \Vert d_z \Vert \\
    &\leq -\mu\vert\xi\vert_{\attr}^2 + 2\Vert P(T_2)\Vert \cdot \left( \frac{\kappa}{2}\Vert z \Vert^2 + \frac{1}{2\kappa} \Vert d_z \Vert^2 \right) \\
    &\leq -\mu\vert\xi\vert_{\attr}^2 + \kappa\Vert P(T_2)\Vert\Vert z \Vert^2 + \frac{\Vert P(T_2)\Vert}{\kappa} \Vert d_z \Vert^2 \\
    &\leq -\left( \mu - \kappa\Vert P(T_2)\Vert\right)\vert\xi\vert_{\attr}^2 + \frac{\Vert P(T_2)\Vert}{\kappa} \Vert d_z \Vert^2 \\
    &\leq -\frac{(\mu - \kappa\Vert P(T_2)\Vert)}{\alpha_2}V(\xi) + \frac{\Vert P(T_2)\Vert}{\kappa} \delta_{\max}^2.
\end{aligned}
\end{equation}
Note, $d_z\in P_z$, $P_z$ being compact, and $\delta_{\max}= \sup \{\Vert d \Vert\colon d\in P_z\}$ imply $\Vert d_z \Vert\leq \delta_{\max}$.
Also, since $\kappa\in(0,\mu/\Vert P(T_2)\Vert)$, one has that $\mu-\kappa\Vert P(T_2)\Vert >0$.

When $\xi\in D$, $\xi$ evolves according to $\xi^+ \in G(\xi)$, where the jump map $G$ is defined in~\eqref{eqn: jump map}.
Moreover, for $\xi\in D$ and $g\in G(\xi)$, the change in $V(\xi)$
is given by $\Delta V(\xi) = V(g)-V(\xi)$.
Without loss of generality, suppose $\tau_q=0$ for some $q\in\verts$.
It then follows that
\begin{equation} \label{eqn: V Jump}
\begin{aligned}
    \Delta V(\xi) &= (z^+)^\top P(\tau^+)(z^+) - z^\top P(\tau) z \\
    &= -\tilde{\vartheta}_q^\top P_{2,q}\exp(\sigma\tau_q)\tilde{\vartheta}_q \leq 0.
\end{aligned}
\end{equation}
Thus, $V(\xi)$ is non-increasing during jumps.

Next, fix a maximal solution $\phi$ of $\hyb$, select $(t,j)\in\dom\phi$, and let $0= t_0 \leq t_1 \leq ... \leq t_j\leq t$ satisfy
\begin{equation*}
    \dom\phi\bigcap([0,t_j]\times\{ 0,1,...,j-1\})  =\bigcup_{k=1}^j ([t_{k-1},t_k]\times\{k-1\}).
\end{equation*}
For every $k\in [j]$ and for almost all $h\in I_{k-1}\coloneqq [t_{k-1},t_k]$, such that the interior of $I_{k-1}$ is nonempty (i.e., $I_{k-1}$ is non-degenerate), one has $\phi(h,k-1)\in C$.
Moreover, since $V(\xi)$ is continuously differentiable and $\phi$ is absolutely continuous, $V\circ \phi$ is absolutely continuous, and, thus, differentiable almost everywhere, over each non-degenerate $I_{k-1}$. 
For every $k\in [j]$ and for almost all $h\in I_{k-1}$ such that $I_{k-1}$ is non-degenerate, \eqref{eqn: generalized time derivative} and~\eqref{eqn: VDot 2} imply
\begin{equation} \label{eqn: V differential inequality}
\begin{aligned}
    &\frac{\dd}{\dd h} V(\phi(h,k-1)) \leq \dot{V}(\phi(h,k-1)) \\
    &\leq  -\frac{(\mu - \kappa\Vert P(T_2)\Vert)}{\alpha_2}V(\phi(h,k-1)) + \frac{\Vert P(T_2)\Vert}{\kappa} \delta_{\max}^2.
\end{aligned}
\end{equation}
The integration of both sides of~\eqref{eqn: V differential inequality} over a non-degenerate $I_k$ leads to 
\begin{equation} \label{eqn: V solution over flow interval}
\begin{aligned}
    V(\phi(t_k,k-1)) &\leq V(\phi(t_{k-1},k-1))\exp\left(-\bar{\mu}(t_k-t_{k-1})\right) \\
    &\phantom{=} \ + \frac{\Vert P(T_2)\Vert \delta_{\max}^2}{\bar{\mu}\kappa}
\end{aligned}
\end{equation}
for each appropriate $k\in [j]$. 
Note, $\bar{\mu}$ is defined in the statement of Theorem~\ref{thm: A is GPES}.
Similarly, for each $k\in [j]$
with $\phi(t_k,k-1)\in D$, \eqref{eqn: V Jump} implies
\begin{equation} \label{eqn: V jump along solutions}
    V(\phi(t_k,k)) \leq V(\phi(t_k,k-1)).
\end{equation}
By inductively stitching the inequalities in~\eqref{eqn: V solution over flow interval} and~\eqref{eqn: V jump along solutions} along the maximal solution $\phi$, it follows that
\begin{equation} \label{eqn: V GES Bound}
    V(\phi(t,j)) \leq V(\phi(0,0))\exp\left(-\bar{\mu}t\right) + \frac{\Vert P(T_2)\Vert \delta_{\max}^2}{\bar{\mu}\kappa}.
\end{equation}
Using the left inequality in~\eqref{eqn: (t,j) inequality} and the identity $t = \varepsilon t + (1-\varepsilon)t$, one can derive
\begin{equation} \label{eqn: t+j inequality}
    -\bar{\mu}t \leq -\min\left\{ \bar{\mu}\varepsilon,\frac{\bar{\mu}(1-\varepsilon)T_{\min}}{N}\right\} (t+j) + \bar{\mu}(1-\varepsilon) T_{\min},
\end{equation}
which, when substituted into~\eqref{eqn: V GES Bound}, yields
\begin{equation} \label{eqn: V GES Bound 2}
\begin{aligned}
    V(\phi(t,j)) &\leq V(\phi(0,0))\exp\big(\bar{\mu}(1-\varepsilon) T_{\min}\big) \\
    &\phantom{=} \ \cdot\exp\left(-\min\left\{ \bar{\mu}\varepsilon,\frac{\bar{\mu}(1-\varepsilon)T_{\min}}{N}\right\} (t+j)\right) \\
    &\phantom{=} \ + \frac{\Vert P(T_2)\Vert \delta_{\max}^2}{\bar{\mu}\kappa}.
\end{aligned}
\end{equation}
The application of the inequalities in~\eqref{eqn: V Bounds} on~\eqref{eqn: V GES Bound 2} yield
\begin{equation} \label{eqn: V GES Bound 3}
\begin{aligned}
    \vert\phi(t,j)\vert_{\attr}^2 &\leq \frac{\alpha_2}{\alpha_1}\vert\phi(0,0)\vert_{\attr}^2 \exp\big(\bar{\mu}(1-\varepsilon) T_{\min}\big) \\
    &\phantom{=} \ \cdot \exp\left(-\min\left\{ \bar{\mu}\varepsilon,\frac{\bar{\mu}(1-\varepsilon)T_{\min}}{N}\right\} (t+j)\right) \\
    &\phantom{=} \ + \frac{\Vert P(T_2)\Vert \delta_{\max}^2}{\alpha_1\bar{\mu}\kappa},
\end{aligned}
\end{equation}
which leads to the desired bound in~\eqref{eqn: GPES bound}.
Hence, $\attr$ is GPES.
Note, if $t+j \geq T$ such that $(t,j)\in\dom\phi$, the right-hand side of~\eqref{eqn: GPES bound} and~\eqref{eqn: nu time} imply $\kappa_1\exp(-\alpha(t+j))\vert \phi(0,0)\vert_{\attr} + \kappa_2 \leq \nu$.
Thus, $\hyb$ achieves $\nu$-approximate synchronization by~\eqref{eqn: GPES bound}.

By~\eqref{eqn: V} and~\eqref{eqn: V GES Bound 2}, $\phi_z(t,j)$ is bounded for all $(t,j)\in\dom\phi$.
Since $\WV$ is full rank and $\phi_z(t,j)$ is bounded, $\phi_{\vartheta}(t,j)$, $\phi_{\tilde{\vartheta}}(t,j)$, $\phi_{\tilde{a}}(t,j)$, and $\phi_{\tilde{\theta}}(t,j)$ are bounded for all $(t,j)\in\dom\phi$.
Since $a_p$ is constant for all $p\in\verts$, $\phi_{\vartheta}(t,j)$ is bounded, and $\phi_{\tilde{\vartheta}}(t,j)$ is bounded, \eqref{eqn: agent controller}--\eqref{eqn: aTilde_p} can be used to show $\phi_{u_p}(t,j)$ is bounded for all $(t,j)\in\dom\phi$.
\end{proof}

In Theorem~\ref{thm: A is GPES}, the GPES of $\attr$ for $\hyb$ requires $\mmat(\tau)\prec 0$ (be negative definite) for all $\tau\in\tspc$, and such a sufficient condition cannot be verified in practice as there are an infinite number of points in $\tspc$ to check.
Ergo, a practical means of validating the negative definiteness of $\mmat(\tau)$ over $\tspc$ is required, which motivates the following result
\begin{corollary} \label{cor: LMI Relaxation}
    Let $\widetilde{\fmat}$ denote the orthogonal complement of $\fmat$.
    If $\widetilde{\fmat}Q(0_N)\widetilde{\fmat}^{\top} \prec 0$, then $\mmat(\tau)\prec 0$ for all $\tau\in\tspc$.
\end{corollary}
\begin{proof}
    Suppose the hypothesis, and fix a $\tau\in\tspc$. 
    By employing~\eqref{eqn: theorem 1 items}, one can see that
    \begin{equation} \label{eqn: Alternative MMat}
        \widetilde{\fmat}^{\top}\mmat(\tau)\widetilde{\fmat} = \widetilde{\fmat}^{\top}(\fmat^\top P(\tau) + P(\tau)\fmat + Q(\tau))\widetilde{\fmat} = \widetilde{\fmat}^{\top}Q(\tau)\widetilde{\fmat}.
    \end{equation}
    Furthermore, $Q(\tau)\preceq Q(0_N)$, that is $Q(0_N) - Q(\tau)$ is positive semi-definite, since $Q(\tau)$ is a diagonal matrix with diagonal elements that are non-increasing functions of $\tau$.
    Observe that the rank of $\fmat$ is $3N-1$ by construction, which implies $\widetilde{\fmat}$ has full rank by the Rank theorem.
    Therefore, 
    \begin{equation} \label{eqn: Q Bound}
        \widetilde{\fmat}^{\top}Q(\tau)\widetilde{\fmat} \preceq \widetilde{\fmat}^{\top}Q(0_N)\widetilde{\fmat}.
    \end{equation}
    By the hypothesis, $\widetilde{\fmat}$ being full rank, \eqref{eqn: Alternative MMat}, and~\eqref{eqn: Q Bound}, $\mmat(\tau)\prec 0$.
    Since $\tau$ was arbitrary, the desired result follows.
\end{proof}
In light of Corollary~\ref{cor: LMI Relaxation}, one needs to only satisfy $\mmat(\tau)\prec 0$ at a single point in $\tspc$, namely, $\tau=0_N$, and, if the hypothesis of Corollary~\ref{cor: LMI Relaxation} is satisfied, then we can guarantee the $\mmat(\tau)\prec 0$ holds for all $\tau\in\tspc$.

\section{Simulation Example} \label{sec: Numerical}
\begin{figure}
\centering 
\includegraphics[width=0.9\columnwidth]{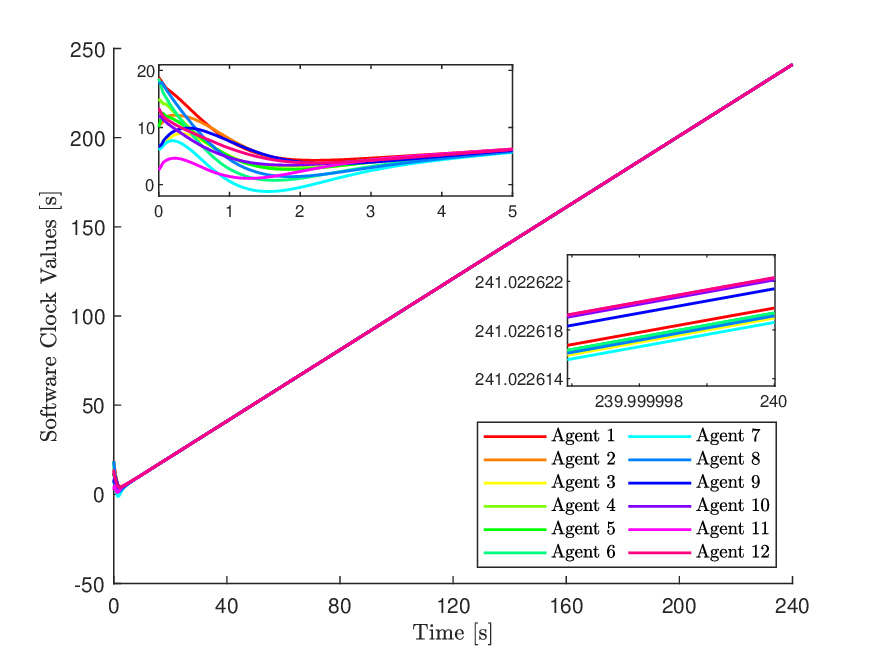} 
\caption{Depiction of the trajectories of the software-defined times, $\{\phi_{\vartheta_p}(t,j)\}_{p\in\verts}$.
The left inset plot shows the software-defined time trajectories during the beginning of the simulation, while the right inset plot shows the software-defined time trajectories during the end of the simulation.}
\label{fig: Software Times}
\vspace{-5pt}
\end{figure}

\begin{figure}
\centering 
\includegraphics[width=0.9\columnwidth]{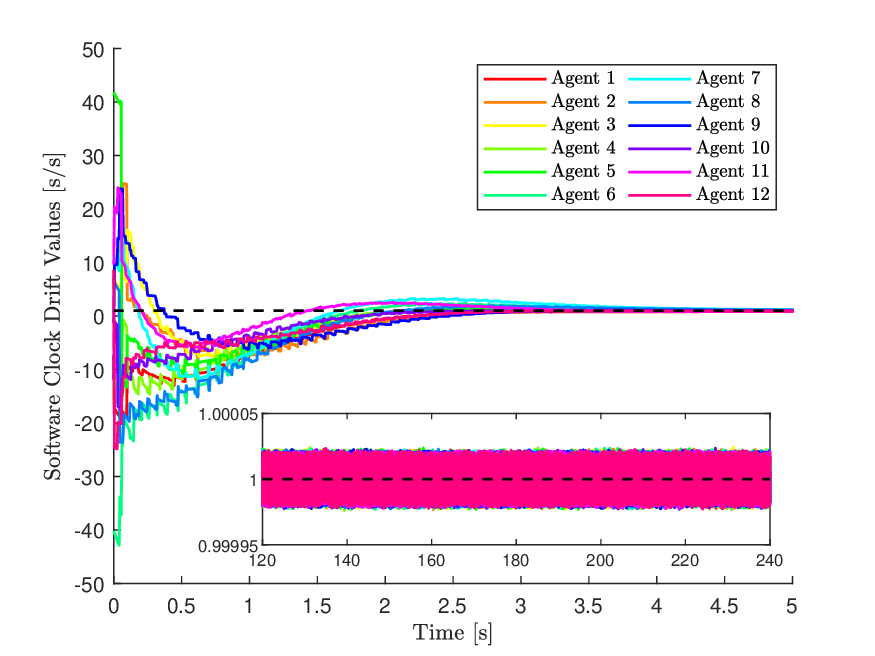} 
\caption{Illustration of the trajectories of the software-defined time drifts, i.e., $\{\phi_{\dot{\vartheta}_p}(t,j)\}_{p\in\verts}$.
The main plot shows the drift trajectories during the beginning of the simulation; the inset plot shows the drift trajectories during the second half of the simulation.
The black dashed line is the graph of the function $t\mapsto a^{\star}=1$, representing the desired drift, in both plots.
For each agent of the MAS, the drift trajectory converges to $[1-\epsilon,1+\epsilon]$ with $\epsilon = 2.27\times 10^{-5}$.}
\label{fig: Software Clock Drift Values}
\vspace{-5pt}
\end{figure}

\begin{figure}
\centering 
\includegraphics[width=0.9\columnwidth]{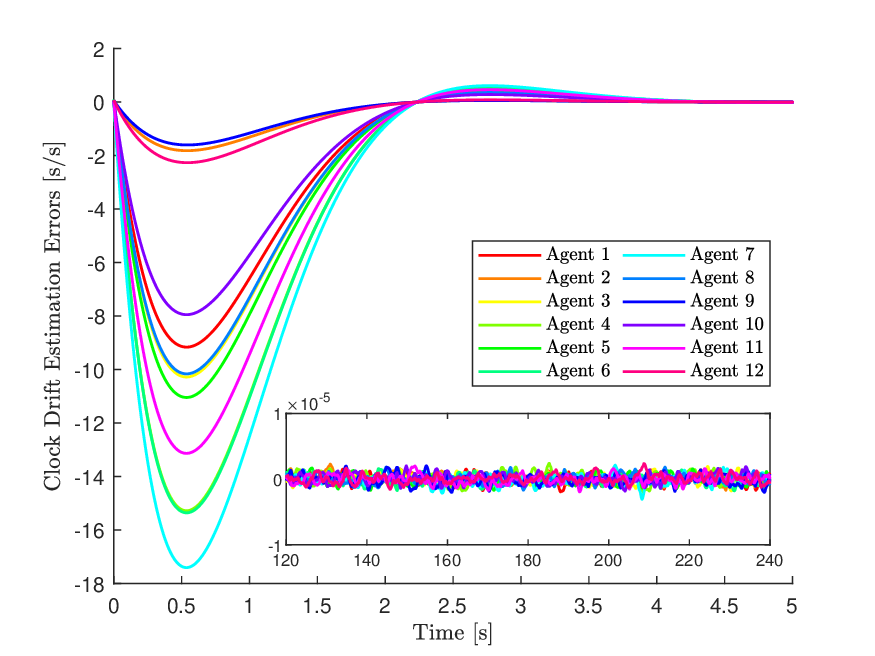} 
\caption{Illustration of the trajectories of the hardware clock drift estimation error, i.e., $\{\phi_{\tilde{a}_p}(t,j)\}_{p\in\verts}$
The main plots depicts the drift estimation error trajectories during the beginning of the simulation; the inset plots shows the drift estimation error trajectories during the second half of the simulation.
For each agent of the MAS, the drift estimation error trajectory converges to $[-\epsilon,\epsilon]$ with $\epsilon = 3.06\times 10^{-6}$.
}
\label{fig: Drift Estimation Errors}
\end{figure}

\begin{figure}
\centering 
\includegraphics[width=0.9\columnwidth]{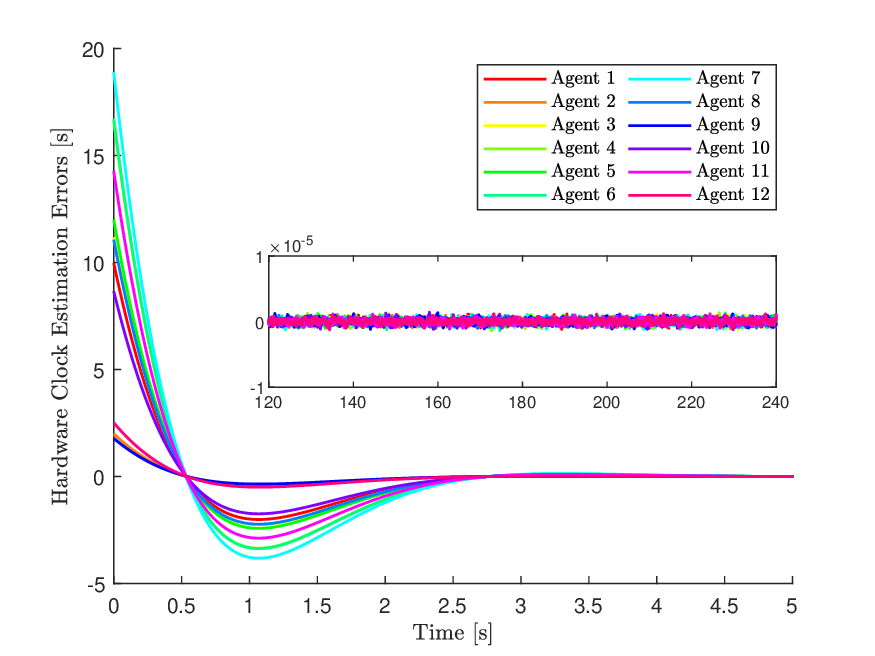} 
\caption{Depiction of the trajectories of the hardware clock estimation errors, $\{\phi_{\tilde{\theta}_p}(t,j)\}_{p\in\verts}$.
The main plot shows the hardware clock estimation error trajectories during the beginning of the simulation; the inset plot shows the hardware clock estimation error trajectories during the second half of the simulation.
For each agent of the MAS, the hardware clock estimation error trajectory converges to the set $[-\epsilon,\epsilon]$ with $\epsilon = 1.18\times 10^{-6}$.
}
\label{fig: Hardware Clock Estimation Errors}
\end{figure}

\begin{figure}
\centering 
\includegraphics[width=0.9\columnwidth]{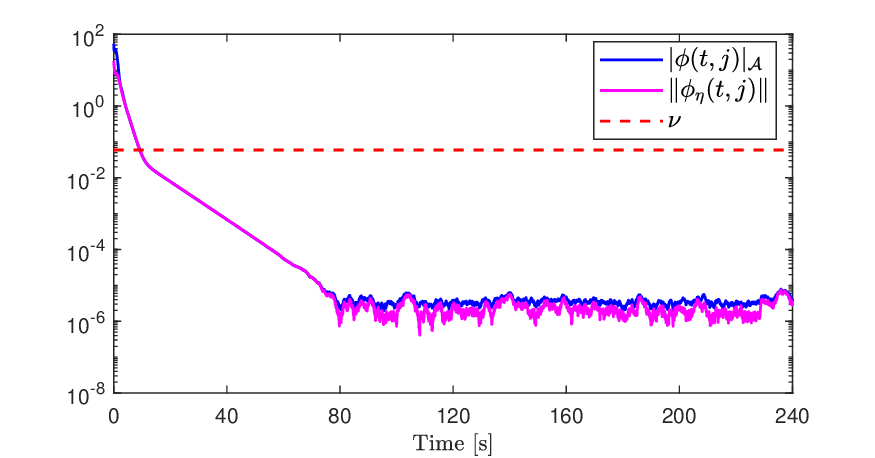} 
\caption{The blue line depicts the trajectory of the distance between the solution $\phi$ of the closed-loop hybrid system $\hyb$ and the set $\attr$.
The magenta line represents the trajectory of the disagreement metric $\Vert\eta\Vert$ along the solution $\phi$.
The red dashed line is the graph of the function $t\mapsto \nu=0.06$, representing the desired time synchronization tolerance.
The horizontal axis uses a linear scale, and the vertical axis uses a logarithmic scale.
Both trajectories are bounded above by $8\times 10^{-6}$ for $t\geq 80$ seconds.}
\label{fig: Distance to Attractor}
\vspace{-5pt}
\end{figure}

\begin{figure}
\centering 
\includegraphics[width=0.9\columnwidth]{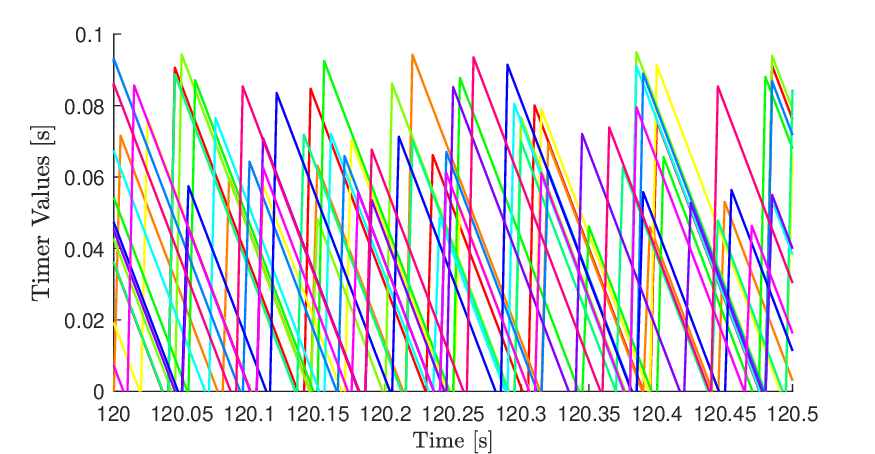} 
\caption{Depiction of the trajectories of the software-defined timers, $\{\phi_{\tau_p}(t,j)\}_{p\in\verts}$.
Due to the small size of the parameters $T_1^p,T_2^p$ for each $p\in\verts$ relative to the length of the simulation, we showcase the timer trajectories over the interval $[120,120.5]$.
We omit the legend, but the same coloring of the trajectories used in Figures~\ref{fig: Software Times}--\ref{fig: Hardware Clock Estimation Errors} apply to this figure.}
\label{fig: Timer Values}
\end{figure}

In this section, we present results gathered from a numerical simulation that implements the~\texttt{ChronoSync} algorithm, i.e., the items in~\eqref{eqn: hardware clock p}--\eqref{eqn: agent controller} for each agent.
The simulation parameters are $n=2$, $N=12$, $\nu= 0.06$, $k_u=0.72$, $k_a=4.2$, $k_{\theta}=3$, $a^{\star}=1$, $\delta_p = 20$ parts per million for every $p\in\verts$, $\sigma = 35$, and $T_1^p=0.05$ seconds and $T_2^p=0.1$ seconds for all $p\in\verts$.
Due to limited space, we omit the adjacency matrix $\adj$, $P_1$, $P_2(T_2)$, and $P_3$.
Nevertheless, we do report that the Fiedler value of the Laplacian $\lap$ corresponding to $\graph$ is $\lambda_2(\lap) = 0.167$.
Further, $\alpha_1 =3.214$, $\alpha_2 = 140.742 = \Vert P(T_2)\Vert$, $\mu = 0.656$, $\kappa = 0.002$, $\bar{\mu} = 0.102$, $\delta_{\max} = 9.79\times 10^{-5}$, and $\kappa_2 = 0.042$.
The matrices $P_1$, $P_2(0_N)$, and $P_3$ that define the block diagonal matrix $P(0_N)$ can be used to show that $\mmat(0_N)$ is a symmetric, negative definite matrix and that the hypothesis of Corollary~\ref{cor: LMI Relaxation} is satisfied.
The simulation results are provided in Figures~\ref{fig: Software Times}--\ref{fig: Timer Values}.
As evidenced by Figure~\ref{fig: Distance to Attractor}, the \texttt{ChronoSync} algorithm was able to achieve the specified synchronization tolerance of $\nu= 0.06$.

\section{Conclusion}
This work develops \texttt{ChronoSync}, a novel decentralized time synchronization protocol for MASs based on consensus dynamics.
Despite the presence of bounded disturbances and desynchronized time kept by hardware clocks, \texttt{ChronoSync} enables all agents to steer their software-defined clocks into practical synchronization, achieve a common user-defined drift in the software clocks, and accurately estimate the unknown hardware clock drifts.
Furthermore, the degree of practical synchronization is commensurate with the net perturbation bound.
By design, \texttt{ChronoSync} supports directed, intermittent, and asynchronous communication between agents.
Although not discussed in this work, \texttt{ChronoSync} can readily be extended  to accommodate switching coupling graphs that evolve in a piecewise constant manner.

In the future, one can relax the assumption that information broadcasts are received instantaneously and simultaneously by all neighboring agents.
This can be done by accounting for communication delays, which are present when communicating beyond the local horizon or in space, and packet dropouts, which tend to our when multiple agents broadcast information at the same time over the same frequency band.
Hence, it is of particular interest to the authors to account for the physics of communication and efficiently use allocated frequency bands through techniques like Code Division Multiple Access.
\bibliographystyle{IEEEtran}
\bibliography{References}

\end{document}